# Forecasting the U.S. Real House Price Index

Vasilios Plakandaras[+], Rangan Gupta*, Periklis Gogas[+♦] and Theophilos Papadimitriou[+],

[+] Department of Economics, Democritus University of Thrace, Greece.

*Department of Economics, Pretoria University, South Africa.

**Abstract**

The 2006 sudden and immense downturn in U.S. House Prices sparked the 2007 global financial crisis and revived the interest about forecasting such imminent threats for economic stability. In this paper we propose a novel hybrid forecasting methodology that combines the Ensemble Empirical Mode Decomposition (EEMD) from the field of signal processing with the Support Vector Regression (SVR) methodology that originates from machine learning. We test the forecasting ability of the proposed model against a Random Walk (RW) model, a Bayesian Autoregressive and a Bayesian Vector Autoregressive model. The proposed methodology outperforms all the competing models with half the error of the RW model with and without drift in out-of-sample forecasting. Finally, we argue that this new methodology can be used as an early warning system for forecasting sudden house prices drops with direct policy implications.

---

♦ Corresponding author: Periklis Gogas, Email: pgkogkas@ierd.duth.gr, tel. +302531039555,

## 1.     Introduction and literature review

The last global financial crisis restored a keen interest in both academic and policy circles on the role of asset prices and in particular housing prices in economic activity. As Leamer (2007) notes the housing market predicted eight of the ten post World War II recessions, acting as a leading indicator for the real sector of the economy. In fact he goes as far as to state that "Housing is the Business cycle". Vargas and Silva (2008) argue that housing prices adjustments play an important role in the determination of the phase of the business cycle. When the economy booms, construction and employment in the housing sector expand rapidly to respond to excess demand, rapidly pushing nominal house prices upwards. During the contraction phase, the drop in private income reduces aggregate demand and nominal house prices. Nevertheless, nominal house prices typically fall sluggishly since householders are unwilling to lower their prices. Most of the adjustment is achieved through decreases in sales volume resulting in a drop in the construction sector and the housing based employment. Moreover, during contraction and recession real house prices fall rapidly as general inflationary trends reduce real house prices even with sticky nominal prices.

Recently, several authors reach to empirical findings that house prices can be instrumental in forecasting output. (Forni *et al*, 2003; Stock and Watson, 2003; Gupta and Das, 2010; Das *et al*, 2009; 2010; 2011; Gupta and Hartley, 2013). The housing construction sector represents a large part of total economic activity expressed in the GDP. Consequently, as it reflects a large portion of the overall wealth of the economy, house prices fluctuations can be an indicator of the evolution of GDP (Case *et al*, 2005). As it is the case with other assets, the movement of house prices can be also an indicator of the future direction of inflation (Gupta and Kabundi, 2010). Overall, accurate forecasting of the evolution path of house prices can be a useful tool both to house market participants and monetary policy authorities.

There is a vast literature regarding U.S. house prices forecasting. Rapach and Strauss (2007) use an autoregressive distributed lag (ARDL) model framework, containing 25 determinants to forecast real housing price growth for the individual states of the Federal Reserve's Eighth District. They find that ARDL models tend to outperform a benchmark AR model. Rapach and Strauss (2009) extend the same analysis on the 20 largest U.S. states based on ARDL models examining state, regional and national level variables. Once again, the authors reach similar conclusions on the importance of combining forecasts of models with different lag structure. Gogas and Pragidis (2011) use the risk premium calculated as the difference between various long-term interest rates and the agents' expectations about future short-term rates as input variable in predicting the future direction of house prices. They conclude that investors and analysts can use effectively the information provided by the interest rate risk premium today in order to estimate the probability of obtaining a below-trend S&P CS-10 index three months ahead.

Gupta and Das (2010) also forecast the recent downturn in real house price growth rates for the twenty largest U.S. states. The authors use Spatial Bayesian VARs (BVARs), based only on monthly real house price growth rates, to forecast their downturn over the period 2007:01 to 2008:01. They find that BVAR models are well-equipped in forecasting the future direction of real house prices, though they significantly underestimate the decline. They attribute this under-prediction of the BVAR models to the lack of any information on fundamentals in the estimation process.

Das et al., (2010) use small-scale BVARs, Bayesian Factor Augmented VARs (BFAVARs) and large-scale BVARs in forecasting house prices of the nine census regions. The authors use the standard Minnesota Bayesian prior in estimating the Bayesian models. They indicate that the BFAVARs are best-suited in forecasting the house price growth rates of the nine census divisions. Gupta et al (2011) examine the explanatory power of small and large sets of economic variables, using atheoretical models such as VAR, BVAR, FAVAR, BVAR, BFAVAR, and forward-looking structural Dynamic Stochastic General Equilibrium (DSGE) models. Based on the average root mean-squared errors for the one-, two-, three-, and four-quarters-ahead forecasts, they find that the small-scale Bayesian-shrinkage model fed with 10 variables outperforms all the other input sets. Gupta (2013) uses dynamic factor and Bayesian shrinkage models in a large number of predictors (145 variables) and forecasts house prices for four U.S. census regions and for the aggregate economy. The results show that the BFAVAR models exhibit the best forecasting ability. Similar results were also obtained by Gupta and Kabundi (2010) and Aye and Gupta (forthcoming) using Bayesian predictive regressions in forecasting the overall US house price index. More recently, Balcilar et al., (forthcoming) compared the ability of nonlinear AR models in forecasting nominal house price growth rates of the four U.S. census regions and the aggregate economy, relative to an AR model. Interestingly, even though they could detect nonlinearity in the in-sample for all the 5 growth rates of house prices, when it came to out-of-sample point, interval and density forecasting, the evidence in favor of the nonlinear model was virtually non-existent[1]. In light of the above discussion, it is clear that, in general, Bayesian models are best-suited in forecasting house prices - regional or for the aggregate US.

In this paper we build on previous empirical studies by comparing the econometric Bayesian Vector Autoregressive (BVAR) and Bayesian Autoregressive (BAR) models, instead of Bayesian predictive regressions to avoid issues of endogeneity, with a novel forecasting methodology on one-year-ahead forecasting. We propose a methodology that combines Ensemble Empirical Mode Decomposition from the field of signal processing with the machine learning Support Vector Regression methodology for constructing forecasting models. In comparison to previous studies we use a much longer sample spanning annual observations from 1890 to 2012 and evaluate the use of 11 macroeconomic variables. Finally, we expand our research framework into multi-period ahead forecasting in an effort to evaluate our forecasting methodology as a house prices early-warning system, focusing on the 2006-2007 house prices downturn.

The rest of the paper is organized as follows: methodology is discussed in section 2, dataset is presented in section 3 while the empirical results are reported in section 4. Finally, section 5 concludes.

## 2. Methodology

### 2.1 Overview

A common issue among econometricians when developing forecasting models is that almost all time series are aggregations of more volatile elements. For instance, the annual dataset of real annual housing prices in the U.S. that consists of quarterly or even monthly price surveys which are aggregated. As the sampling frequency decreases short-term dynamics diminish and long-term characteristics such as trend or seasonality are to be observed. The use of such input variables in a forecasting system incorporates error, making difficult to create accurate models. Smoothing techniques can reduce the influence of noise and errors in observed data, offering a less volatile representation of the underlying phenomenon. Nonetheless, essential information may be lost during the smoothing process.

---
[1]For a detailed literature review on forecasting involving the U.S. commercial and residential real-estate markets, refer to Ghysels et al., (2013).

A key issue in all smoothing methodologies is the definition of the optimal point, beyond which smoothing results to distortion rather than noise reduction. In real applications, noise-free time series are unobservable and we can only approximate them through their smoothed counterparts. Many smoothing implementations select *ad hoc* the smoothness degree imposed; see for example the Hodrick-Prescott (1997) filter and the discussion regarding the optimum value of the smoothness parameter $\lambda$ with respect to data frequency. To mitigate proper model selection dilemmas, our implementation exploits a relatively novel signal decomposition method called Ensemble Empirical Mode Decomposition (EEMD) used as a smoothing function. The main advantages of this smoothing framework are a) the absence of *a priori* assumptions and that b) the tuning of the model's parameters is solely based on data characteristics.

In this paper we propose a three step forecasting methodology as follows:

Step 1. A smoothing of the initial series employing the EEMD
Step 2. A variable selection process using the Elastic Net approach
Step 3. Fitting a Support Vector Regression model for forecasting.

In the first step we decompose all input series with EEMD into a low frequency (smoothed) and a remaining highly volatile (fluctuating) component that approximate the long term trend and short term dynamics[2], respectively. By de-composing the initial time series into two distinct components we are able to use a distinct forecasting model for each of the individual components. In this way each model is better suited to the individual data characteristics. The model regarding the smoothed component of the output time series is fed with the smoothed component of the input time series. The same mechanism is used accordingly for the fluctuating component. The variable selection in every case is conducted using the Elastic Net (Zou and Hastie, 2005) and the selected variable set is forwarded to an SVR model. The proposed system yields one forecasted time series for the smoothed and one for the fluctuating component of the initial time series. Overfitting is avoided by adopting a 4-fold Cross Validation scheme in the training step. The sum of the two forecasted components is the final output of our methodology and it is evaluated for its out-of-sample forecasting accuracy. An overview of the method is depicted in Figure 1.

---
[2] For more information about the representation of short and long term data characteristics during EMD decomposition see Wu *et al* (2007).

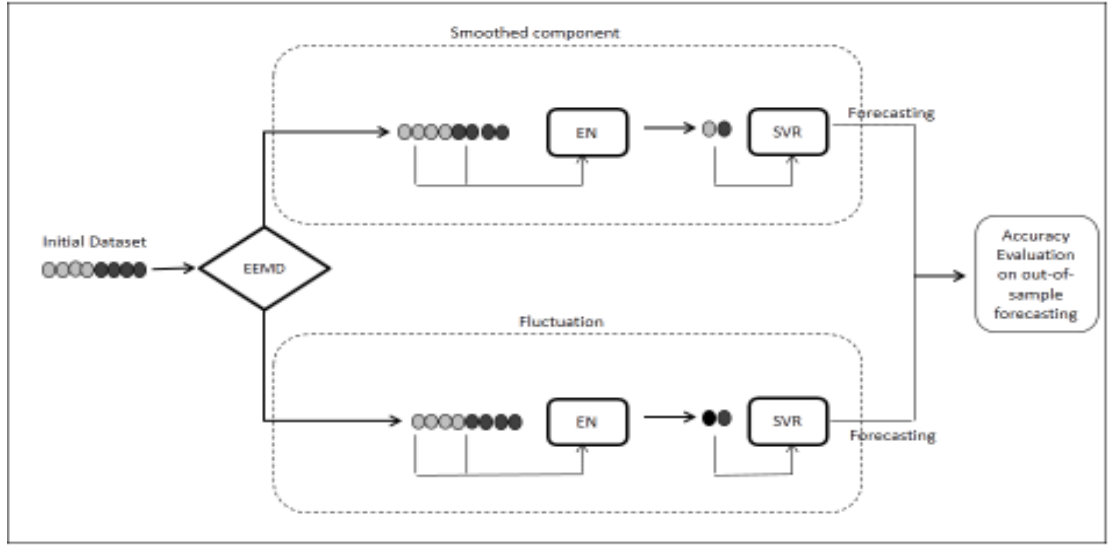

**Figure 1:** The hybrid EEMD-EN-SVR structural model.

## 2.2  VAR and BVAR Specifications

Following the work of Sims (1980), we create an unrestricted VAR model as follows:

$$\mathbf{y}_t = A_o + A(L)\mathbf{y}_t + e_t, \qquad (1)$$

assuming $\varepsilon_t \sim N(0, \sigma^2 I_n)$ and $I_n$ is an $n \times n$ identity matrix where $\mathbf{y}_t$ is an $n \times 1$ vector of variables, $A(L)$ is an $n \times n$ polynomial matrix in the backshift operator L with lag length p [ $A(L) = A_0 + A_1 L + A_2 L^2 + \cdots + A_p L^p$ ], $A_0$ is an $n \times n$ vector of constant terms, and $e_t$ is an $n \times n$ vector of error terms.

VAR models use equal lag lengths for all variables in the model, which implies that the researcher must estimate many parameters, including some that may be statistically insignificant. This over-parameterization problem can create multicollinearity, leading to possibly large out-of-sample forecasting errors. Litterman (1981, 1986), Doan *et al,* (1984) and Spencer (1993) use a Bayesian VAR (BVAR) model to overcome the over-parameterization problem. Rather than eliminating statistically insignificant coefficients, they impose different weights on the coefficients across different lag lengths, assuming that longer lags coefficients are less significant than shorter lag ones. In our Bayesian variants of the classical VAR and VEC models we follow the propositions of Litterman (1981) that proposes a diffuse prior to the constant, often named in literature as the "Minnesota prior". Formally, the means and variances of the Minnesota prior take the form:

$$\beta_i \sim N\left(1, \sigma^2_{\beta_i}\right) \text{ and } \beta_j \sim N\left(0, \sigma^2_{\beta_j}\right) \qquad (2)$$

where $\beta_i$ are the coefficients associated with the lagged dependent variables in each equation of the VAR model and $\beta_j$ are all other coefficients. Setting all variances to zero, the prior specification for every variable reduces to a Random Walk model with a drift. The prior variances, $\sigma^2_{\beta_i}$ and $\sigma^2_{\beta_j}$, specify uncertainty about the prior with means $\overline{\beta_\iota} = 1$, and $\overline{\beta_\iota} = 0$, respectively.

Doan *et al.* (1984) propose that standard deviations could be calculated based on a small numbers of hyper-parameters: *q, g*, and a weighting matrix $f(i,j)$. The standard deviation of the prior imposed on variable *j* in equation *i* at lag *m*, for all *i, j* and *m* is as follows:

$$S(i,j,m) = [q \times g(m) \times f(i,j)]\frac{\hat{\sigma}_\iota}{\hat{\sigma}_j}, \qquad (3)$$

$$\text{Subject to} \begin{cases} f(i,j) = 1, \text{if } i = j \text{ and } k_{ij} \text{ otherwise, with } 0 \leq k_{ij} \leq 1 \\ g(m) = m^{-d}, \text{with } d > 0 \end{cases}$$

where $\hat{\sigma}_\iota$ is the estimated standard error of the univariate autoregression for variable *i*. The ratio $\frac{\hat{\sigma}_\iota}{\hat{\sigma}_j}$ scales the variables to account for differences in the units of measurement, while term *q* indicates the overall tightness of the model, with the prior getting tighter as the value falls. The parameter $g(m)$ measures the tightness on lag *m* with respect to the first lag and tightens the prior as the lag order increases. The weighting matrix $f(i,j)$ equals the tightness of variable *j* in equation *i* and by increasing the interaction

(i.e., the value of $k_{ij}$), we loosen the prior.[3] Sims *et al.* (1990) argue that with the Bayesian approach entirely based on the likelihood function, the associated inference does not require stationarity, since the likelihood function exhibits the same Gaussian shape regardless of the presence of unit roots. Thus, in this paper we specify BVAR and BAR models in levels.

### *2.3 Ensemble Empirical Mode Decomposition and time series smoothing*

The EEMD is a data driven algorithm that decomposes a time series into finite additive oscillatory components called Intrinsic Mode Functions (IMFs). Proposed by Wu and Huang (2009), the decomposition into IMFs is achieved through an iterative scheme until the residual has no local extrema. The amplitude of the added Gaussian noise during the decomposition procedure is chosen according to the maximized relative RMSE criterion proposed by Guo and Tse (2013)[4].

An example of a decomposition applied on the real annual US House Prices is depicted in Figure 2.

---

[3] For an illustration, see Dua and Ray (1995).
[4] For more details on the method see Wu and Huang (2009)

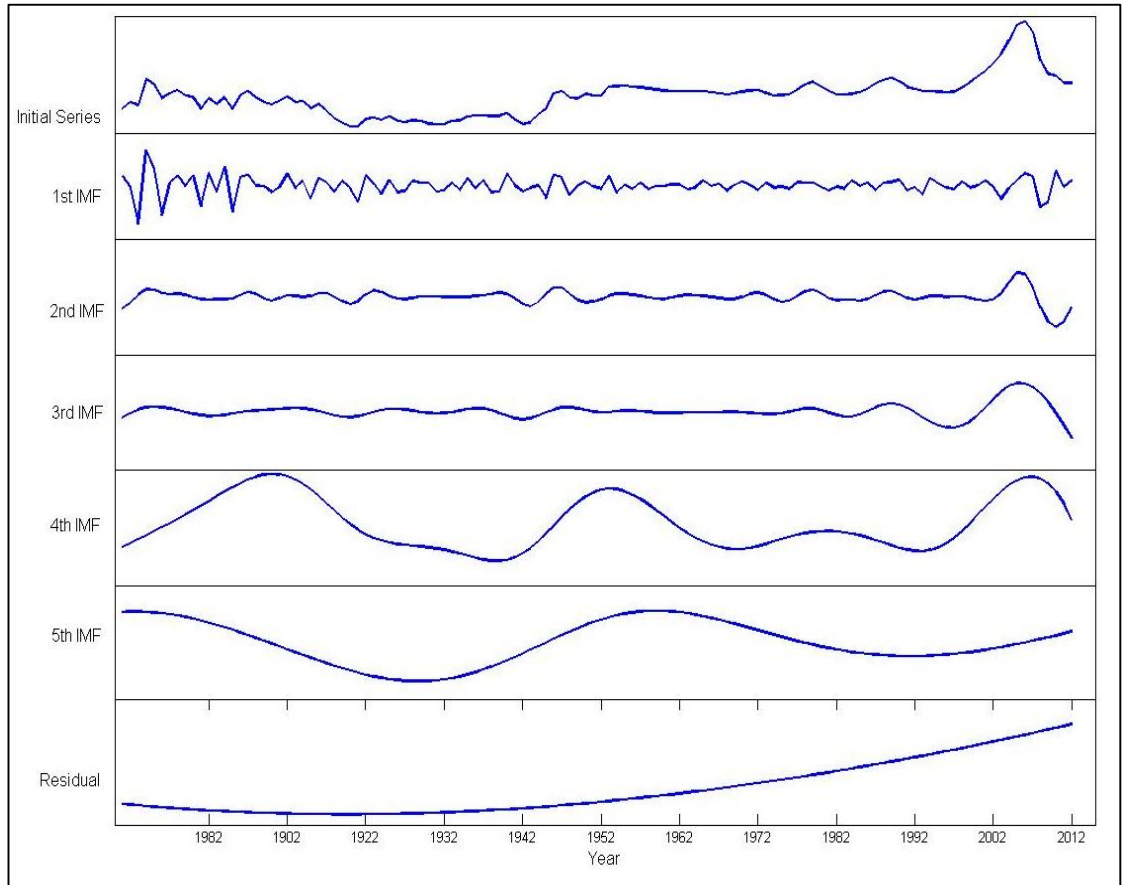

**Figure 2**: Decomposition of the original (1st row) real annual US House Prices into 6 series. The last row (straight line) is the residue of the EEMD process.

During decomposition, the frequency of every IMF drops as its index number increases. Wu et al (2007) propose that each IMF represents different dynamics of the time series, with the more volatile IMFs depicting short-run dynamics and the less volatile ones long-run trends of the phenomenon. Moghtaderi *et al.* (2013) build on this framework proposing a trend extraction technique based on decomposition. They argue that long run trends should be examined as summations rather than independent IMFs, since individual characteristics are dispersed between IMFs and are not exclusively isolated in only one IMF. In other words they suggest that the long-run trend is the summation of 1) the lower frequency IMFs and 2) the final residual of the EEMD decomposition.

Expanding the proposition of Moghtaderi *et al* (2013) we argue that all IMFs can be evaluated for constructing a representation of short-run dynamics of the initial timeseries and a long-run trend function. The former is the result of the summation of more volatile IMFs while the latter can be obtained following the procedure described by Moghtaderi *et al* (2013). Thus, the segregation problem breaks down to selecting the most appropriate IMF index which defines the limit between short and long term dynamics. With mathematical notation, the above smoothing function is expressed with:

$$\boldsymbol{h}_{i*} = \sum_{i=i_*}^{R} \boldsymbol{IMF}^i + r, \qquad 1 \leq i \leq L \qquad (4)$$

where $\boldsymbol{h}_{i*}$ is the smoothed variant of the initial series for index $i_*$, R the total number of IMFs, $r$ the final EEMD residual and $i_*$ the index of the IMF beyond which we start the summation. For instance, in Figure 3 the decomposition of the real house prices time series results into 5 IMFs and a final residual. By summing the 3$^{rd}$, 4$^{th}$ and 5$^{th}$ IMF with the final residual, we obtain the smoothed function of the red curve of Figure 3.

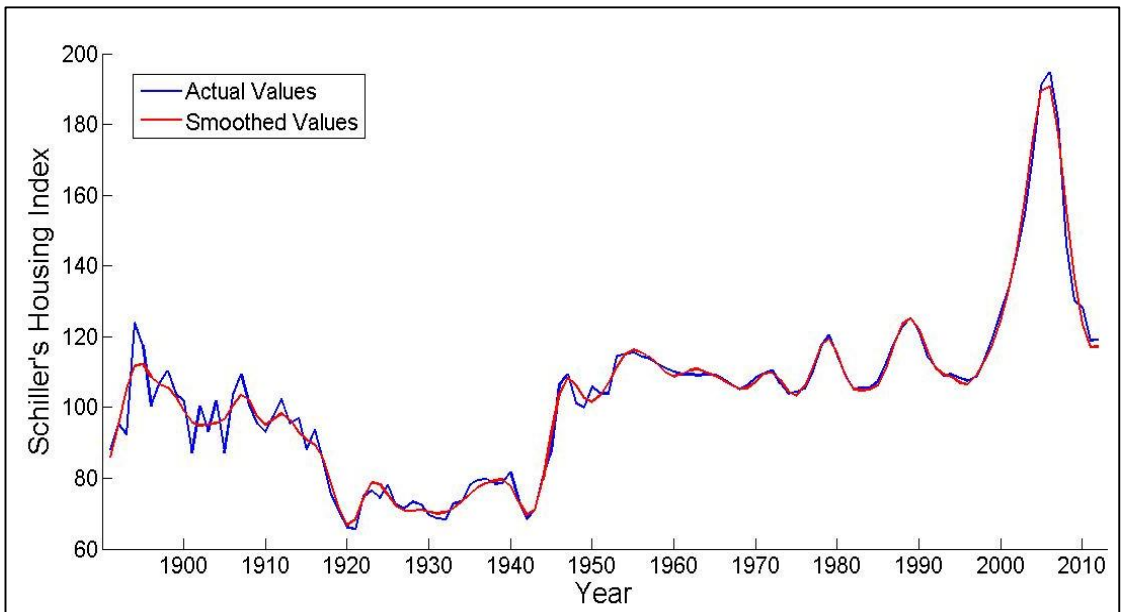

**Figure 3:** Real U.S. House Prices and its EEMD smoothed series.

## 2.4 Support Vector Regression (SVR)

The Support Vector Regression is a direct extension of the classic Support Vector Machines algorithm proposed by Vladimir Vapnik (1992) and originates from the field of statistical learning. Despite the ability of the methodology to provide accurate forecasts with high generalization ability, it has not attracted significant interest in forecasting economics and financial time series. Among the few empirical applications, Rubio *et al* (2011) forecast electric loads based on SVR models, while Papadimitriou *et al* (forthcoming) develop SVM models for directional price forecasting in electric energy markets. Härdle *et al* (2009) evaluate the default risk of companies with SVM and Öğüt *et al* (2012) extend default risk forecasting in the banking sector. Khandani *et al* (2010) use SVR models for private credit risk evaluation and Papadimitriou *et al* (forthcoming) use SVR models for recession forecasting. Finally, Gogas *et al* (2013) compare Simple Sum and Divisia monetary aggregates under a machine learning framework in order to forecast the U.S. GDP.

The basic idea is to find a linear function that has at most a predetermined deviation from the actual values of the dataset. In other words, we do not care about the error of each forecast as long as it doesn't violate the predefined threshold, but we penalize any deviations higher than the threshold. The set that bounds this "error-tolerance band" is the Support Vector (SV). This, is located through a minimization procedure.

One of the main advantages of the SVR in comparison to other machine learning techniques is that, it typically yields a convex minimization problem, avoiding local minima (Vapnik, 1992). The model is built in two steps: the training and the testing step. In the training step, the largest part of the dataset is used for the estimation of the function (i.e. the detection of the Support Vectors that define the "error-tolerance band"); in the testing step, the generalization ability of the model is evaluated by checking the model's performance in the small subset that was left aside during training.

Using mathematical notation and starting from a training dataset $D = [(x_1, y_1), (x_2, y_2), \ldots (x_n, y_n)], x_i \in \mathbb{R}^m, y_i \in \mathbb{R}, i = 1, 2, \ldots n$, where for each observation pair, $x_i$ are the observation samples and $y_i$ is the dependent variable (the target of the regression system) the linear regression function takes the form of $f(x) = w^T x + b$. The SVR methodology tries to reach two contradictory goals: a) find a solution that best approximates the given dataset (i.e. a large part of the datapoints should be inside the tolerance "belt", while a few points will lie out of bounds) and b) to find a solution that generalizes to the underlying population. This is achieved by solving:

$$\min \left( \frac{1}{2} \|w\|^2 + C \sum_{i=1}^{n} (\zeta_i + \zeta_i^*) \right) \qquad (5)$$

$$\text{subject to} \begin{cases} y_i - (\boldsymbol{w}\boldsymbol{x}_i + b) \leq \varepsilon + \zeta_i \\ (\boldsymbol{w}\boldsymbol{x}_i + b) - y_i \leq \varepsilon + \zeta_i^* \\ \zeta_i, \zeta_i^* \geq 0 \end{cases}$$

where ε defines the tolerance belt around the regression, and $\zeta_i, \zeta_i^*$ are slack variables controlled through a penalty parameter C (see Figure 4). All the points inside the tolerance belt have $\zeta_i, \zeta_i^* = 0$. The problem (5) is a convex quadratic optimization problem with linear constraints and has a unique solution. The first part of the objective function controls the generalization ability of the regression, by imposing the smaller possible $\|\boldsymbol{w}\|$. This is not an obvious statement and a detailed analysis of the SVR minimization process is not in the scope of this paper, however we can hint that the smaller is $\|\boldsymbol{w}\|$, the closer to parallel to the x-axes is the regression function. Geometrically we can see that a parallel line to the x-axes, maximizes the covered area by the tolerance belt, which means maximum generalization ability. The second part of the objective function controls the regression approximation to the training data points (by increasing C we penalize with a bigger weight any point outside the tolerance belt i.e. with $\zeta_i \geq 0$ or $\zeta_i^* \geq 0$). The key element in the SVR concept is to find the balance between the two parts in the objective function, controlled by the ε and C parameters, as presented in Figure 4.

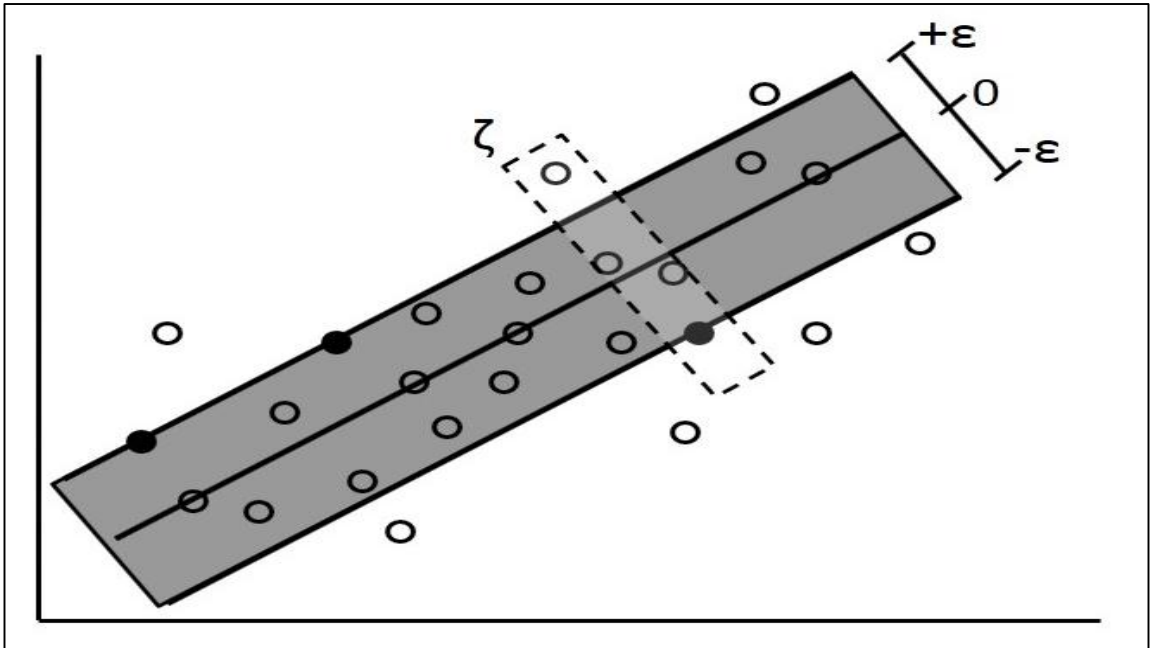

**Figure 4:** Upper and lower threshold on error tolerance indicated with letter ε. The boundaries of the error tolerance band are defined by the Support Vectors (SVs) denoted with the black filled points. Forecasted values greater than ε get a penalty ζ according to their distance from the tolerance accepted band.

and the solution is given by:

$$\boldsymbol{w} = \sum_{i=1}^{n}(a_i - a_i^*)\boldsymbol{x}_i \qquad (6)$$

and
$$y = \sum_{i=1}^{n}(a_i - a_i^*)\boldsymbol{x}_i^T \boldsymbol{x} \qquad (7)$$

Real life phenomena are rarely described correctly by a linear regression; they are too complex for such a simplistic approximation. An alternative method to treat real phenomena datasets would be to project them into a higher dimensional space where the transformed dataset may be described by a linear function. The "kernel trick" follows the projection idea while ensuring minimum computational cost: the dataset is mapped in an inner product space, where the projection is performed using only dot products within the original space through special "kernel" functions, instead of explicitly computing the mapping of each data point. Non-linear kernel functions have evolved the SVR mechanism to a non-linear regression model.

In our simulations we employed four kernels: the linear, the radial basis function (RBF), the sigmoid and the polynomial. The mathematical representation of each kernel is:

Linear
$$K_1(\boldsymbol{x}_1, \boldsymbol{x}_2) = \boldsymbol{x}_1^T \boldsymbol{x}_2 \qquad (8)$$

RBF
$$K_2(\boldsymbol{x}_1, \boldsymbol{x}_2) = e^{-\gamma\|\boldsymbol{x}_1 - \boldsymbol{x}_2\|^2} \qquad (9)$$

Polynomial
$$K_3(\boldsymbol{x}_1, \boldsymbol{x}_2) = (\gamma \boldsymbol{x}_1^T \boldsymbol{x}_2 + r)^d \qquad (10)$$

Sigmoid
$$K_4(\boldsymbol{x}_1, \boldsymbol{x}_2) = \tanh(\gamma \boldsymbol{x}_1^T \boldsymbol{x}_2 + r) \qquad (11)$$

with factors $d$, $r$, $\gamma$ representing kernel parameters.

## 2.5 Elastic Net

The Elastic Net is a variable selection method proposed by Zou and Hastie (2005) that linearly combines the LASSO (Timmerman, 1996) and the ridge regression techniques. According to the LASSO, when solving a regression problem we try to minimize the squared difference of actual to forecasted values, while imposing the constraint that only the coefficients with an absolute value greater than a threshold are acceptable. Coefficients with a value lower than the threshold are set to zero and thus discarded. In this way the LASSO methodology tries to minimize the forecasting error of the regression model, while pruning regressors. The difference of ridge regression to the LASSO methodology is that instead of imposing a threshold on the absolute value of each coefficient, it evaluates its squared value and discards all coefficients with a squared value below a threshold.

The Elastic Net combines the aforementioned methods in a more flexible framework. Instead of using a fixed form (absolute or squared value) for each coefficient, Zou and Hastie (2005) propose a regularization parameter that fluctuates from zero to unit. When zero is selected, the Elastic Net reverts to the LASSO model, while when the regularization parameter is set to one, we get a ridge regression representation. In other words, the LASSO and the ridge regression are only special cases of an Elastic Net model. Overall, the aforementioned methodology is flexible, adapts to data characteristics and can select regressors that lead to accurate regression models.

## 3     The Dataset

In comparison to previous studies we use a much longer sample spanning annual observations from 1890 to 2012, with an in-sample size of 1890-1988 and an out-of-sample of 1989-2012. A very short sample may describe only episodic trends of the entire phenomenon, while longer periods allow to observe the forecasting performances of these models over prolonged periods and changes in the trend of house prices. The decision to use an out-of-sample starting in 1989 is to cover a period of sharp increase in activity in the housing market (Rapach and Strauss, 2007; 2009). Our dataset consists of ten annual U.S. macroeconomic variables spanning the period 1890 to 2012, and in turn, are selected based on literature:

- ➢ The real GDP per capita (RGDPPC) (Agnello and Schuknecht, 2011; Case and Shiller, 1990)
- ➢ The long term interest rate (LTR) and the short term interest rate (STR) (McGibany and Nourzad, 2004; Mikhed and Zemčík, 2009 Agnello and Schuknecht, 2011)

- The population (POP) (Agnello and Schuknecht, 2011; Mikhed and Zemčík, 2009; Case and Shiller, 1990; Case and Mayer, 1996)
- The real stock price (RSP) (Abelson et al., 2005; Mikhed and Zemčík, 2009; Rapach and Strauss, 2009)
- The real construction cost (RCONSTR)(Case and Shiller, 1990; Jud and Winkler, 2002; Mikhed and Zemčík, 2009; Zeno and Füss, 2010)
- The unemployment rate (UNEMPL) (Case and Mayer, 1996; Abelson et al., 2005; Rapach and Strauss, 2007)
- The inflation rate (INFL) (Stevenson, 2000; Abelson et al., 2005; Rapach and Strauss, 2007)
- The real oil price (ROILP) (Padilla, 2005; Beltratti and Morana, 2010)
- The ratio of budget deficit/surplus as a fiscal policy indicator (FISPOL) (Afonso and Sousa, 2011, 2012; Agnello and Schuknecht, 2011; Agnello and Sousa, forthcoming)

The dataset consists of 11 annual U.S. macroeconomic variables spanning the period 1890 to 2012. This include the real house price (RHP), the fiscal policy variable (FISPOL), real GDP per capita (RGDPPC), unemployment (UNEMPL), long term interest rate (LTR), short term interest rate (STR), inflation rate (INFL), population (POP), real construction cost (RCONSTR), real stock price (RSP) and real oil price (ROILP). All variables are from Robert J. Shiller web page, barring real GDP, population, unemployment and part of the budget surplus/deficit data used for computing the fiscal policy variable which are from the Global Financial data (GFD) base. For the budget surplus/deficit, we obtain the 1890 to 2006 data from GFD while 2007 to 2012 data are obtained from the Federal Reserve Bank, St. Louis. We use the ratio of budget surplus/deficit to GDP as our measure of fiscal policy. Inflation rate is computed as annual rate of growth in consumer price index. Real oil price is obtained by deflating the nominal West Texas Intermediate with CPI. Figure 5 depicts real house prices and the grey areas represent recession periods reported by the National Bureau of Economic Research. During 2006 to 2009, the historically highest price levels are followed by a sharp drop in real houses prices of unprecedented magnitude.

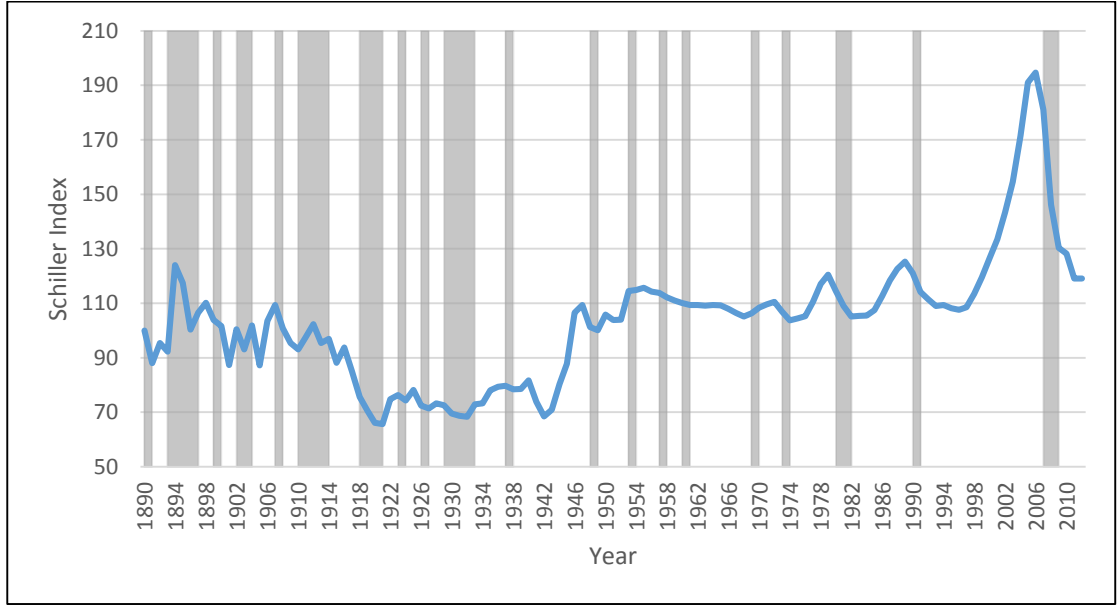

**Figure 5:** Yearly Real Housing Prices in the U.S. as measured by the Schiller housing prices index (1890=100). Grey areas indicate recession periods of the U.S. economy as reported by the National Bureau of Economic Research (NBER).

## 3. *Empirical Results*

In order to test the generalization ability of our selected model, our dataset is split in two parts for in- and out-of-sample forecasting. The ratio chosen is 80/20. The forecasting accuracy is measured by the Mean Absolute Percentage Error (MAPE) and Directional Symmetry (DS):

$$MAPE = \frac{100}{n}\sum_{i=1}^{n}\left|\frac{\hat{y}_i - y_i}{y_i}\right| \qquad (12)$$

$$DS = \frac{100}{n}\sum_{i=1}^{n} d_i, where\ d_i = \begin{cases} 1\ if\ (y_i - y_{i-1})(\hat{y}_i - \hat{y}_{i-1}) > 0 \\ 0\ otherwise \end{cases} \qquad (13)$$

where $\hat{y}_i$ and $y_i$ stand for the forecasted and actual values respectively while $n$ is the total number of out-of-sample observations. The MAPE measures the absolute percentage error in the forecast, while the DS is a measure of directional forecasting accuracy. Directional forecasting is of key interest to house market participants and policy authorities, since as discussed, the future direction of house prices can be an indicator of the current and future phase of the business cycle.

In order to test the forecasting ability of the proposed EEMD-EN-SVR methodology in forecasting the real house prices index, we compare it with several alternative forecasting models. Apart from the EEMD-EN-SVR we develop a Random Walk (RW) and a simple autoregressive version without any extra explanatory variables labeled EEMD-AR-SVR; only the past values of the smoothed and the fluctuating part of the house prices index are used as input variables. The optimum lag structure for both the EEMD-AR-SVR and EEMD-EN-SVR model is selected according to the minimum in-sample MAPE. For the EEMD-AR-SVR model we select two lags for the fluctuating component and six lags for the smoothed one. Following the shrinkage scheme of the BVAR model described in Section 2, we introduce lagged values of all predictors and leave EN to choose among them the most informative ones. In other words as in the BVAR framework the EN may select the second lag of a variable but exclude the first one, on the contrary to classic VAR models. The EN selects one input variable for the fluctuating part and 36 for the smoothed one[5].

Moreover, we also develop an autoregressive BAR and a BVAR model that have been used extensively in house prices forecasting literature. A key advantage of the shrinkage procedure of the BVAR models is that they adapt to data characteristics, excluding irrelevant variables and focusing on the more informative ones with respect to forecasting. The lag structure selection for the Bayesian models is based on the minimum Schwartz Information Criterion (SIC) (Schwartz, 1978) value. For both the BAR and the BVAR model the selected lag is one[6]. We consider various values for the tightness term $q$ the decay factor $g$ and the standard deviation of the distribution of the prior $k_{ij}$. The values $q$=0.3, $g$=0.5 and $k_{ij} = 0.5$ yield the highest in-sample accuracy and they are used to train the BVAR to attain the out-of-sample forecasts. The BAR model is obtained by setting the interaction term $k_{ij} = 0.001$, while q and g remains the same as in the BVAR model[7]. As proposed by the literature, we fix the parameter $a$ of the EN penalty term to 0.5, since simulations with $a = 0.1, 0.3, 0.7$ and $0.9$ yield quantitative similar results.

---

[5] The selected variables can be accessed at
http://utopia.duth.gr/~vplakand/Selected_Variables_by_the_Elastic_Net.pdf
[6] The application of the Akaike Information Criterion (Akaike, 1974) and Hannah-Quinn Information Criterion (Hannah and Quinn, 1979) as expected selected higher lag order. We follow the SIC that results in more parsimonious models.
7 As suggested by the extant Bayesian literature, we also experimented with the following combinations of q and g respectively: (0.2, 1.0); (0.1, 1.0); (0.2, 2.0); (0.1, 2.0).

In Table 1 we report the in-sample and out-of-sample forecasting accuracy in one period ahead static forecasting.

**Table 1: In and out-of-sample forecasting results**

|  | In-sample accuracy | | Out-of-sample accuracy | |
| --- | --- | --- | --- | --- |
| Model | MAPE (%) | DS (%) | MAPE (%) | DS (%) |
| RW | 4.732 | 56.701 | 5.352 | 70.833 |
| RW with a drift | 6.588 | 52.577 | 5.387 | 70.833 |
| BAR | 4.952 | 55.670 | 5.422 | 70.833 |
| BVAR | 4.809 | 56.701 | 11.931 | 75.000 |
| EEMD-AR-SVR(linear) | 1.985* | 84.615* | 2.151* | 87.500* |
| EEMD-EN-SVR(linear) | 2.337 | 78.495 | 5.990 | 87.500* |

Note: Best values are noted with an *. Best kernel for the SVR models is noted in parenthesis.

We observe that the highest in-sample accuracy according to the MAPE is achieved with the EEMD-AR-SVR model closely followed by the EEMD-EN-SVR. The latter is also the most accurate model according to the Directional Symmetry criterion. When we evaluate the out-of-sample forecasting ability of our models in the last two columns of Table 1, we observe that the EEMD-AR-SVR model outperforms all other models for both the MAPE and the DS criteria. The performance of the EEMD-EN-SVR drops significantly in terms of the MAPE (the forecasting error doubles) and it is now outperformed by all other models with the exception of the BVAR. In terms of the DS criterion the EEMD-EN-SVR model provides the second most accurate directional forecasting after the EEMD-AR-SVR. Finally, both the BAR and BVAR models perform poorly and exhibit a lower forecasting accuracy both in- and out-of-sample as compared to the SVR models with the exception of the BAR that provides a better MAPE than the EEMD-EN-SVR out-of-sample. Moreover, both the BAR and BVAR models provide less accurate forecasts than the simple naïve Random Walk model in-sample and out-of-sample. On the other hand, since the autoregressive EEMD-AR-SVR model outperforms the RW model, we find evidence against the Efficient Market

Hypothesis (EMH) of Eugene Fama (1965). Since with the EEMD-AR-SVR model the real house price index can be adequately forecasted with only historical (lagged) values of itself we can conclude that even weak form efficiency is not supported by the data in the U.S. housing market.

In order to have a visual representation of the forecasting ability of the models presented in Table 1, we depict in Figure 6 the out-of-sample forecasts.

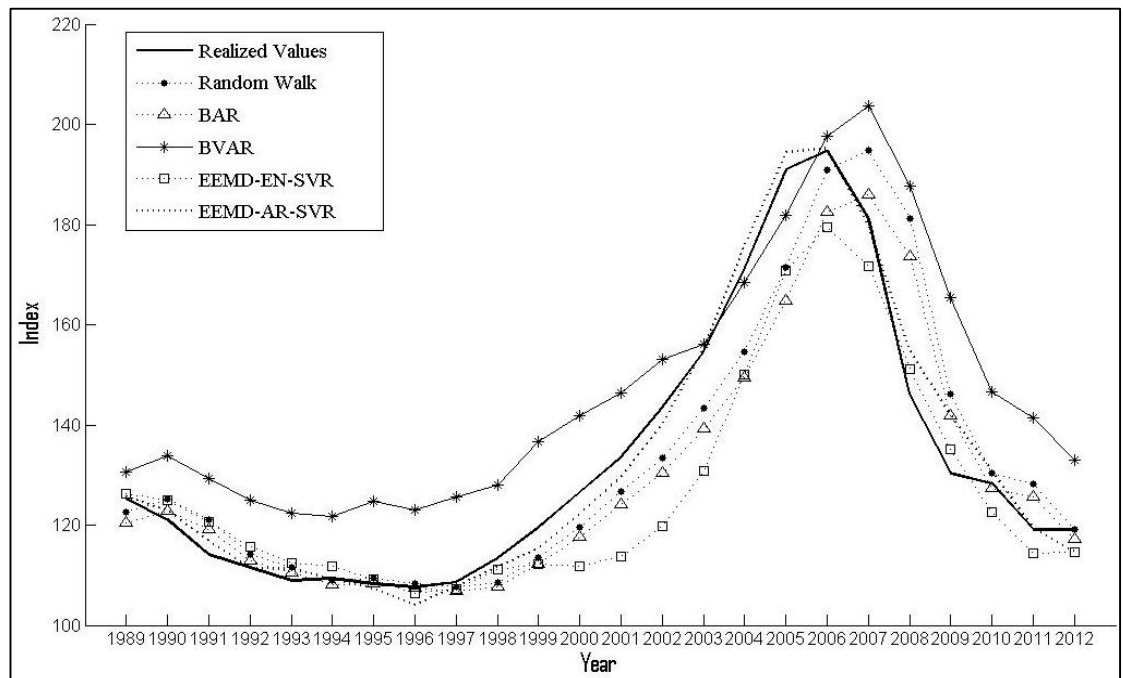

**Figure 6**: Out-of-sample forecasted values

As we observe from Figure 6, the EEMD-AR-SVR model is the only one of the five models tested that timely and efficiently forecasts the 2006-2009 sudden downturn in real U.S. Housing Prices that sparked the latest financial crisis. The EEMD-EN-SVR model performs very poorly with the out-of-sample data. This may be an indication of overfitting for the EEMD-EN-SVR. The other three models, capture the downturn in real house prices with one lag, only one year later than it actually occurred.

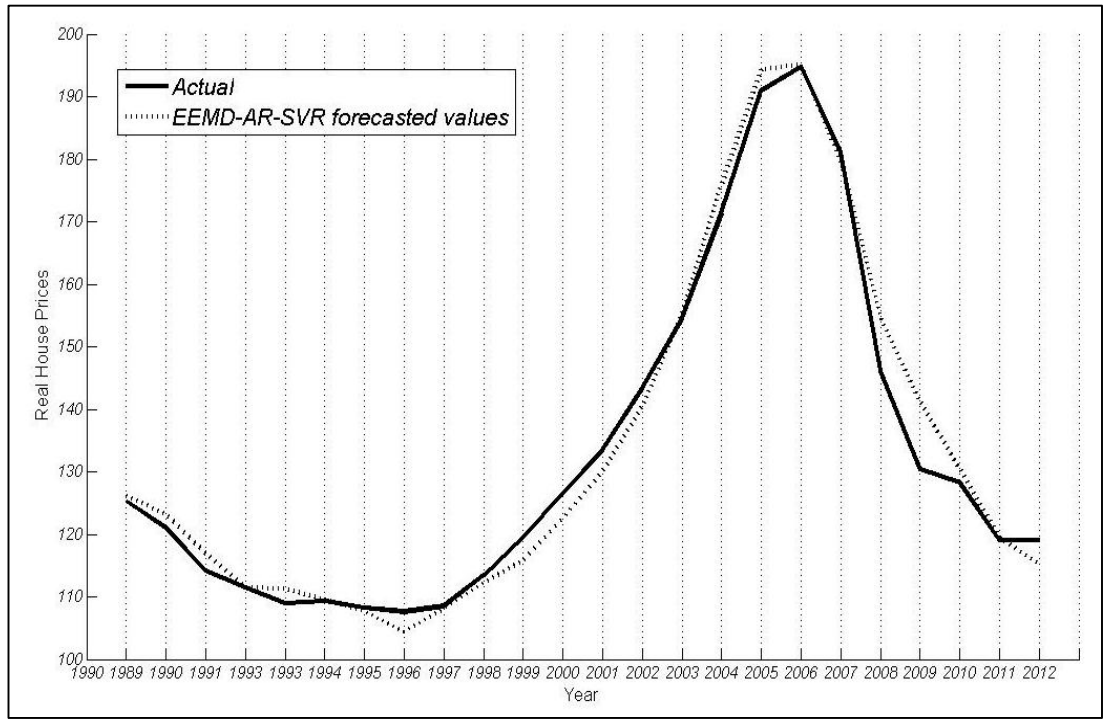

**Figure 7:** EEMD-AR-SVR Out-of-sample forecasted and actual values

In Figure 7 we depict only the actual values and the EEMD-AR-SVR model forecasts. It is clear that the proposed methodology traces very closely the actual prices on the U.S. housing market.

Next, we produce forecasts for longer forecasting windows. From one period (year) ahead that we used thus far, we now extend our forecasting up to ten periods (years) ahead. In this analysis we only use the EEMD-AR-SVR model that outperformed all other tested models in the one period ahead out-of-sample forecasts. The RW model is also included as a benchmark.

| | **Table 2: Multi Period Ahead Forecasting results** | | | | | | | |
|---|---|---|---|---|---|---|---|---|
| | In-Sample forecasting | | | | Out-of-Sample forecasting | | | |
| Period | RW | | EEMD-AR-SVR | | RW | | EEMD-AR-SVR | |
| | MAPE(%) | DS(%) | MAPE(%) | DS(%) | MAPE(%) | DS(%) | MAPE(%) | DS(%) |
| 1 | 4.389 | 55.670 | 1.985* | 84.615* | 5.351 | 70.833 | 2.016* | 87.500* |
| 2 | 6.688 | 52.577 | 3.481* | 71.111* | 10.462 | 58.333 | 4.164* | 83.333* |
| 3 | 8.353 | 45.833 | 5.669* | 50.000* | 14.903 | 54.167 | 9.895* | 62.500* |

| | | | | | | | |
|---|---|---|---|---|---|---|---|
| 4 | 8.664 | 40.000 | 6.766* | 42.046* | 18.672 | 41.667 | 15.344* | 45.833* |
| 5 | 9.667 | 46.809* | 7.987* | 42.529 | 20.735 | 37.500 | 16.528* | 41.667* |
| 6 | 10.232 | 48.387* | 8.180* | 48.148 | 20.760 | 33.333 | 20.566* | 37.500* |
| 7 | 10.640 | 54.348* | 8.655* | 50.633 | 19.874* | 25.000 | 21.630 | 29.167* |
| 8 | 10.594 | 56.044* | 9.419* | 55.844 | 18.133* | 29.167 | 21.085 | 37.500* |
| 9 | 11.201 | 55.556 | 9.757* | 56.000* | 16.364* | 33.333 | 20.300 | 41.667* |
| 10 | 11.426 | 57.303* | 10.210* | 55.405 | 14.713* | 37.500* | 18.049 | 29.167 |

Note: Best forecasts for each forecasting horizon are noted with an *.

As expected, the forecasting accuracy of the model deteriorates as the forecasting horizon increases. Both in in- and out-of-sample forecasting and for both the MAPE and DS criteria the EEMD-AR-SVR model outperforms the RW model for forecasting windows from 1 to 4 years ahead. This superiority of the EEMD-AR-SVR model is maintained in out-of-sample forecasting for both criteria up to 6 years ahead. Moreover, according to the DS the EEMD-AR-SVR outperforms the RW model up to 9 years ahead. Only for the last year (10) of out-of-sample forecasts the RW model is superior to the EEMD-AR-SVR according to both criteria used. For the MAPE and in-sample forecasting the EEMD-AR-SVR is superior to the RW model for all 10 forecasting windows. In order to better observe the forecasting ability of the proposed methodology, we depict the forecasted values of the EEMD-AR-SVR model for the first four horizons in the out-of-sample period in Figure 7.

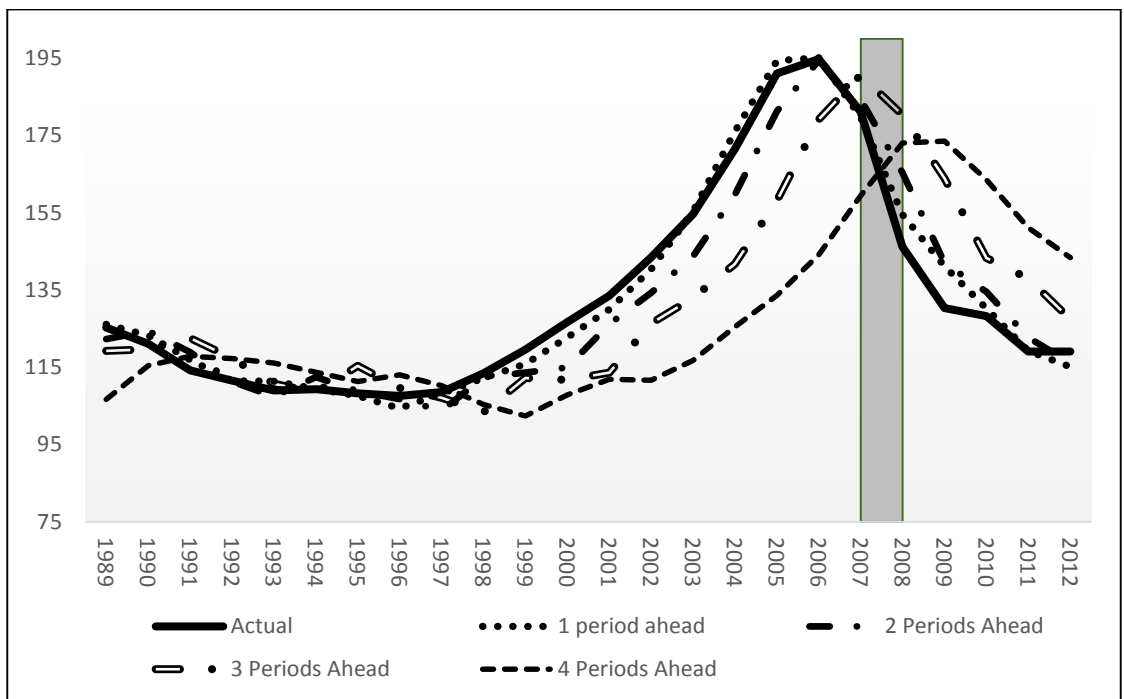

**Figure 8**: Multi period ahead forecasts. Grey areas denoted recession periods of the U.S. economy as reported by the National Bureau of Economic Research.

The model forecasts the significant fall in 2006-2009 of U.S. housing market prices up to 2 periods (years) ahead. Overall, the proposed methodology forecasted correctly the U.S. housing prices collapse nearly two years before the actual event occurred an observation that according to Leamer (2007) can be evaluated as an indicator of the business cycle phase of the economy. As observed in Figure 7, the drop in house prices starting in 2006 signals a contraction period of the U.S. economy followed by the 2007-2008 recession period. The same phenomenon is repeated just before the brief 2 quarters 1991 recession where dropping prices during 1989-1990 mark a contraction period. In other words predicting house prices deterioration can be an indicator that the economy is entering a contraction period and thus monetary policy should be adjusted accordingly.

## *4.    Conclusion*

We examine eleven predictors in forecasting the real U.S. House Prices index comparing the econometric BAR and BVAR techniques and a novel EEMD-SVR forecasting methodology. The empirical findings suggest that the EEMD-AR-SVR outperforms all competing models in both in-sample and out-of-sample forecasting. The reported research framework is extended to single and multi-period ahead forecasting reporting early accurate detection of the 2006-2009 price downturn. Overall, we argue that the proposed forecasting methodology can be used effectively as a policy instrument for determining the business cycle of the economy.

*Acknowledgements*

Vasilios Plakandaras, Dr. Papadimitriou Theophilos and Dr. Periklis Gogas were partly financed in this research by the European Union (European Social Fund – ESF) and Greek national funds through the Operational Program "Education and Lifelong Learning" of the National Strategic Reference Framework (NSRF) - Research Funding Program: **THALES** (MIS 380292). Investing in knowledge society through the European Social Fund.

*References*

Abelson, P., Joyeux, R., Milunovich, G. and Chung, D. (2005) Explaining house prices in Australia: 1970–2003. *The Economic Record*, 81 (S1), S96-S103.

Afonso, A. and Sousa, R.M. (2012). The macroeconomic effects of fiscal policy. *Applied Economics*, 44, 4439–54.

Afonso, A., and Sousa, R.M. (2011). What are the effects of fiscal policy on asset markets? *Economic Modelling*, 28, 1871–90.

Agnello L. and Schuknecht, L. (2011) Booms and busts in housing markets: Determinants and implications. *Journal of Housing Economics* vol. 20, pp.171–190.

Agnello, L., and Sousa, R. M., (Forthcoming). Fiscal policy and asset prices. *Bulletin of Economic Research.*

Balcilar M., Gupta R. and Miller M.S. (forthcoming). The Out-of-Sample Forecasting Performance of Non-Linear Models of Regional Housing Prices in the US." Journal of Real Estate Research.

Beltratti, A. and C. Morana (2010) International house prices and macroeconomic fluctuations. *Journal of Banking & Finance*, 34(3), 533–45.

Bernanke, B., and Gertler, M. (1995). Inside the black box: the credit channel of monetary transmission. *Journal of Economic Perspectives*, 9(4), 27–48.

Case, K.E. and Mayer, C. J. (1996) Regional Housing price dynamics within a metropolitan area Science and Urban Economics 26, 387-407.

Case, K.E. and Shiller, R.J. (1990) Forecasting prices and excess returns in the housing market. *AREUEA Journal* vol.8, 253–273.

Chang C.-C. and Lin C.-J. (2011) LIBSVM: a library for support vector machines. *ACM Transactions on Intelligent Systems and Technology*, vol. 2 (27), pp. 1-27. Software available at http://www.csie.ntu.edu.tw/~cjlin/libsvm.

Cortes C. and Vapnik V. (1995) Support-Vector Networks, *Machine Leaming*, vol 20, pp. 273-297.

Cu Jirong, Zhu Mingcang and Jiang Liuguangyan (2011), Housing proce forecasting based on genetic algorithm and Support Vector Machines, *Expert Systems with Applications*, vol. 38(4), pp. 3383-3386.

Das, S., Gupta, R. and Kabundi, A. (2009). Could we have predicted the recent downturn in the South African housing market? *Journal of Housing Economics*, 18 (4), 325–335.

Das, S., Gupta, R. and Kabundi, A. (2010). The blessing of dimensionality in forecasting real house price growth in the nine census divisions of the US. *Journal of Housing Research*, 19 (1), 89–109.

Das, S., Gupta, R. and Kabundi, A. (2011). Forecasting regional house price inflation: a comparison between dynamic factor models and vector autoregressive models. *Journal of Forecasting*, 30 (2), 288–302.

Dickey, D. and Fuller, W. (1981), "Likelihood Ratio Statistics for Autoregressive Time Series with a Unit Root", *Econometrica*, Vol. 49, 1057-1072.

Doan, T. A., Litterman, R. B., and Sims, C. A. (1984) Forecasting and Conditional Projections Using Realistic Prior Distributions. *Econometric Reviews* vol. 3, pp. 1-100.

Engle, R. and Granger, C. (1987), "Cointegration and Error Correction: Representation, Estimation and Testing", *Econometrica, vol.*55, pp. 251-294.

Fama, E (1965) The Behavior of Stock Market Prices, *Journal of Business*. vol. 38, pp. 34–105.

Forni M., Hallin, M., Lippi, M. and Reichlin, L. (2003). Do financial variables help forecasting inflation and real activity in the euro area? *Journal of Monetary Economics.*

Friedman, J., R. Tibshirani, and T. Hastie (2010) Regularization paths for generalized linear models via coordinate descent. *Journal of Statistical Software*, vol 33(1).

Gogas P. and Pragidis I. (2011) Does the Interest Risk Premium Predict Housing Prices? *International Journal of Economic Research*, vol. 8(1), pp. 1-10.

Gogas P., Papadimitriou T. and Takli E. (2013) Comparison of simple sum and Divisia monetary aggregates in GDP forecasting: A Support Vector Machines Approach, *Economics Bulletin,* vol. 33, no. 2, pp. 1101-1115.

Guo W. and Tse P. (2013), A novel signal compression method based on optimal ensemble empirical mode decomposition for bearing vibration signals, *Journal of Sound and Vibration*, vol.332, pp. 423-441.

Gupta, R and Kabundi, A. and Miller, S.M. (2011) Forecasting the US real house price index: structural and non-structural models with and without fundamentals. *Economic Modelling*, 28 2013–2021.

Gupta, R. (2013) Forecasting house prices for the four census regions and the aggregate US economy in a data-rich environment, *Applied Economics*, 45:33, 4677-4697.

Gupta, R. and Das, S. (2010). Predicting downturns in the US housing market: a Bayesian approach. *Journal of Real Estate Finance and Economics*, 41 (3), 294–319.

Gupta, R. and Hartley, F. (2013) The role of asset prices in forecasting inflation and output in South Africa. *Journal of Emerging Market Finance*, vol. 12(3), pp. 239-291

Gupta, R. and Kabundi, A., 2010. Forecasting real US house price: principal components versus Bayesian regressions. *International Business and Economics Research Journal*, 9 (7), 141–152.

Ghysels E., Plazzi A., Torous W. and Valkanov R. (2013) Forecasting Real Estate Prices, E., Alberto, W., and R. Valkanov, *Handbook of Economic Forecasting: Vol II*, G. Elliott and A. Timmermann (Eds.), Elsevier.

Hastie, T., R. Tibshirani, and J. Friedman (2008) The Elements of Statistical Learning, 2nd edition. Springer, New York.

Härdle, Wolfgang, Yuh-Jye Lee, Dorothea Schäfer, and Yi-Ren Yeh (2009) Variable selection and oversampling in the use of smooth support vector machines for predicting the default risk of companies, *Journal of Forecasting*, vol.28(6), pp. 512-534.

Hodrick, R.J. and Prescott, E.C. (1997), Postwar US business cycles: an empirical investigation, *Journal of Money, Credit, and Banking*, vol. 29, pp. 1-16.

Iacoviello, M. and Neri, S., 2010. Housing market spillovers: evidence from an estimated DSGE model. *American Economic Journal: Macroeconomics*, 2, 125–164.

International Monetary Fund (2000). World Economic Outlook: Asset Prices and the Business Cycle.

Jud, G. D. and Winkler, D. T. (2002) The dynamics of metropolitan housing prices. *Journal of Real Estate Research*, 23 (1/2), 29-45.

Khandani, Amir E., Adlar J. Kim, and Andrew W. Lo (2010), Consumer credit-risk models via machine-learning algorithms, *Journal of Banking & Finance*, vol. 34(11), pp. 2767-2787.

Korobilis, B. (2013) Hierarchical shrinkage priors for dynamic regressions with many predictors. *International Journal of Forecasting*, 29, 43–59.

Krugman Paul (2013), End this Depression Now, W&W Norton and Company, NY, US.

Kwiatkowski D., Phillips, P. C.B., Schmidt, P. and Shin, Y. (1992), "Testing the Null Hypothesis of Stationarity against the Alternative of a Unit Root, *Journal of Econometrics*, vol. 54:, pp. 159-178.

Leamer, E.E. (2007) Housing is the business cycle. *NBER Working Paper* No. 13428, Retrieved from http://www.nber.org/papers/w13428.

Litterman, R. B., (1981) A Bayesian procedure for forecasting with vector autoregressions. *Working Paper*, Federal Reserve Bank of Minneapolis.

Litterman, R. B., (1986) Forecasting with Bayesian vector autoregressions – Five Years of Experience. *Journal of Business and Economic Statistics*, 4(1), 25-38.

McGibany, J. M. and Nourzad, F. (2004) Do Lower Mortgage Rates Mean Higher Housing Prices? *Applied Economics*, 36 (4), 305-313.

Mikhed, V. and Zemčík, P. (2009) Do house prices reflect fundamentals? Aggregate and panel data evidence. *Journal of Housing Economics*, 18, 140–149.

Moghtaderi A., Flandrin P, and Borgnat P, (2013), Trend filtering via Empirical Mode Decompostition, *Computational Statistics and Data Analysis*, vol. 58, pp. 114-126.

Nguyen N, Cripps A. (2001), Predicting housing value: A comparison of multiple regression analysis and artificial neural network. *Journal of Real Estate Research*, vol. 3(22), pp. 313-336.

Öğüt, Hulisi, M. Mete Doğanay, Nildağ Başak Ceylan, and Ramazan Aktaş (2012), Prediction of bank financial strength ratings: The case of Turkey. *Economic Modelling*, vol. 29(3),pp. 632-640.

Padilla, M.A. (2005) The effects of oil prices and other economic indicators on housing prices in Calgary, Canada. Massachusetts Institute of Technology. Dept. of Architecture.

Papadimitriou, T., Gogas, P., Matthaiou, M., & Chrysanthidou, E. (forthcoming). Yield curve and Recession Forecasting in a Machine Learning Framework. *Computational Economics*.

Papadimitriou T., Gogas P. and Stathakis E. (forthcoming), Forecasting Energy Markets using Support Vector Machines, *Energy Economics*.

Quang D A, Grudnitski G. (1993) A neural network analysis of the effect of age on housing values. *Journal of Real Estate Research*, vol. 2(8), pp. 253-264.

Rapach, D.E. and Strauss, J.K. (2007). Forecasting real housing price growth in the eighth district states. Federal Reserve Bank of St. Louis. *Regional Economic Development,* 3(2), 33–42.

Rilling G., Flandrin, P. and Goncalves, P. (2005) Empirical mode decomposition, fractional Gaussian noise and Hurst exponent estimation. *IEEE International Conference on Acoustics, Speech, and Signal Processing*, pp. 489–492.

Rubio, Ginés, Héctor Pomares, Ignacio Rojas, and Luis Javier Herrera (2011) A heuristic method for parameter selection in LS-SVM: Application to time series prediction, *International Journal of Forecasting*, vol.27(3), pp. 725-739.

Schwarz, Gideon E. (1978). "Estimating the dimension of a model". *Annals of Statistics* 6 (2): 461–464.

Shaaf M, Erfani G. (1996) Air pollution and the housing market: A neural network approach. *International Advances in Economic Research,* vol. 4(2), pp. 484-495.

Sims, C. A. (1980). Macroeconomics and Reality. *Econometrica*, vol. 48(1), 1-48.

Sims, C. A., Stock, J. H., and Watson, M. W., (1990) Inference in Linear Time Series Models with Some Unit Roots. *Econometrica* 58, 113-144.

Spencer, D. E., (1993) Developing a Bayesian vector autoregression model. *International Journal of Forecasting*, 9(3), 407-421

Stevenson, S. (2000) A long term analysis of housing and inflation. *Journal of Housing Economics,* 9 (1-2), 24-39.

Stock, J.H. and Watson, M.W. (2003). Forecasting output and inflation: the role of asset prices. *Journal of Economic Literature*, 41(3), 788-829.

Tibshirani, R. (1996). Regression shrinkage and selection via the lasso. *Journal of Royal Statistics Society B*., Vol. 58, No. 1, pages 267-288).

Topel, R.H. and Rosen, S. (1988). Housing investment in the United States. *Journal of Political Economy*, 96 (4), 718–740.

Vapnik, V., Boser, B. and Guyon, I. (1992), A training algorithm for optimal margin classifiers, *Fifth Annual Workshop on Computational Learning Theory*, Pittsburgh, ACM, pp.144–152.

Vargas-Silva, C. (2008).The effect of monetary policy on housing: a factor augmented approach. *Applied Economics Letters*, 15(10), 749-752.

Wu, Z., and N. E Huang (2009) Ensemble Empirical Mode Decomposition: a noise-assisted data analysis method. *Advances in Adaptive Data Analysis*, vol. 1, No.1, pp. 1-41.

Zeno and Füss (2010) Macroeconomic determinants of international housing markets *Journal of Housing Economics* 19, 38–50.

Zou, H. and Hastie, T. (2005). Regularization and variable selection via the elastic net. *Journal of the Royal Statistical Society*, Series B, 67, 301–320